# Analytical Process Scheduling Optimization for Heterogeneous Multi-core Systems

Chien-Hao Chen[1], Ren-Song Tsay[2]

National Tsing Hua University, Department of Computer Science

{chchen[1], rstsay[2]}@cs.nthu.edu.tw

*Abstract*— In this paper, we propose the first optimum process scheduling algorithm for an increasingly prevalent type of heterogeneous multicore (HEMC) system that combines high-performance big cores and energy-efficient small cores with the same instruction-set architecture (ISA). Existing algorithms are all heuristics-based, and the well-known IPC-driven approach essentially tries to schedule high scaling factor processes on big cores. Our analysis shows that, for optimum solutions, it is also critical to consider placing long running processes on big cores. Tests of SPEC 2006 cases on various big-small core combinations show that our proposed optimum approach is up to 34% faster than the IPC-driven heuristic approach in terms of total workload completion time. The complexity of our algorithm is O($N$log$N$) where $N$ is the number of processes. Therefore, the proposed optimum algorithm is practical for use.

## A. INTRODUCTION

Heterogeneous multicore (HEMC) architecture has been widely adopted for its performance and energy efficiency compared to conventional homogeneous multi-core (HOMC) architecture [1, 2]. There are several HEMC design types. One integrates both CPUs and GPUs or DSPs [3-4] into one system with the support of proper programming models [5] and compilation techniques [6]. Other more popular HEMC designs integrate cores with the same instruction-set architecture (ISA) but different computing power [7]. The idea is to leverage the advantages of high-performance *big* cores and energy-efficient *small* cores and hence achieve the seemingly contradictory requirements for high performance and low power consumption.

Big-small HEMC designs have been proven effective in practice [1, 2] because they do not have the cross-ISA execution issue faced by CPU-GPU HEMC designs. Therefore, this paper focuses on the optimum process scheduling issue for big-small HEMC designs. To simplify our discussion later in this paper, we use HEMC to refer only to big-small HEMC.

A big core in HEMCs is designed to attain better process execution performance than a small core by adopting higher clock rates, having better *ILP* (*Instruction-Level Parallelism*), or having better *MLP* (*Memory-Level Parallelism*). Note that both the total instruction execution time and total memory delay caused by load/store accesses contribute to the total execution time of a process. To improve instruction execution time, designers often adopt a faster clock rate or advanced out-of-order pipeline architecture for ILP, i.e., executing independent instructions concurrently. For shorter memory access delays, some adopt reorder buffer, parallel multiple memory access techniques, etc., for memory-level parallelism (MLP).

In practice, each process owns a unique computation and memory access pattern and hence unveils unique ILP and MLP values when running on a big or small core. If we define the performance scaling factor (SF) as the ratio of the process execution time on a small core to that on a big core, nearly every process exhibits a distinctive SF number. A few works actually suggest that considering the difference of process SF values for process scheduling is the key to effectively utilizing

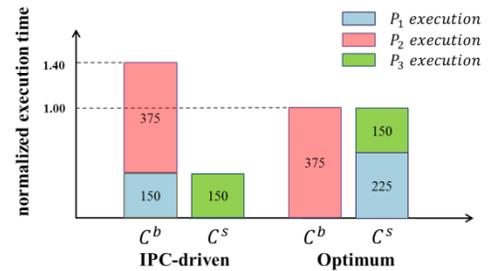

|   | $P_1$ | $P_2$ | $P_3$ |
|---|---|---|---|
| $T^s$ | 225 | 525 | 150 |
| SF | 1.5 | 1.4 | 1.25 |

Table 1: A list of three sample processes with the execution times on a small core, denoted as $T^s$, and the SF values

Figure 1: The normalized completion time comparison of results from IPC-driven approach and our proposed optimum solution method

HEMCs [15, 17], although we shall show later that this idea can sometime be misleading.

In fact, the SF value of a process is not a fixed number. Instead, it fluctuates as the process computation/memory pattern changes dynamically due to program phase behaviors [19]. Consequently, other works propose runtime models [2, 15-20] to predict the SF value of each phase in order to achieve fine-grained, phase-based scheduling optimization. This well-known group of HEMC process scheduling heuristics uses the big-core-best (BCB) method, which attempts to fully utilize big-core computing power. For example, the state-of-the-art IPC-driven approach [15] places processes with larger SFs on big cores and the rest on small cores. This approach usually shortens overall workload completion time by utilizing big cores to accelerate process executions. Nonetheless, none of these heuristics guarantee optimum results.

As illustrated by the example in Figure 1, our proposed optimum scheduling approach can reduce completion time by 40% compared to the IPC-driven approach based on the three sample processes listed in Table 1. In the case under discussion here, we assume each process has one uniform performance phase instead of multiple phases. Thus the execution time on a small core is shortened by a factor of SF times when running on a big core. For instance, in Figure 1 the SF of process $P_1$ is 1.5 and hence its small core execution time of 225 units will become only 150 units on the big core.

This example demonstrates that the IPC-driven approach performs poorly mainly due to the fact that the small core is mostly idling and underutilized. In fact, a process with a long execution time, such as $P_2$, tends to encumber the total completion time if not executed first on the big core. In comparison, the optimum solution actually executes the longest process $P_2$ on the big core and the remaining processes on the small core and produces no idled cores. Clearly, the choice of running the longest process on the big core here contradicts IPC-driven heuristic, which focuses on using the big core for processes with the highest SF.

Thus, we will show that we should consider both the SF value and the process execution time in order to truly optimize solutions. Although we will first show that the scheduling problem can also be formulated as a Linear Programming problem, our proposed exact analytical optimum solution method efficiently solves the problem in $O(N \log N)$ where $N$ is the number of processes. Experimental results show that the optimum results are consistently better than those from other heuristic approaches.

The remainder of this paper is organized as follows. In Section 2, we discuss related work, and in Section 3 we present the analytical optimization method. We then show experimental results in Section 4. In Section 5 we discuss future work and conclude this paper.

## 2. RELATED WORK

The primary goal of HEMC designs is to achieve both high performance and low power consumption although some aim at service quality fairness awareness on HEMCs [8].

In practice, a few HEMC approaches pursue energy efficiency [9, 10] or performance/power efficiency [11-13]. However, most HEMC scheduling works focus on performance optimization [2, 14-20] while utilizing low power cores as much as possible. In this paper, we also tackle the performance optimization problem.

Considering the performance scaling factor (SF) variations for process scheduling is the key to HEMC performance optimization. Some heuristic approaches try to maximize total IPC or IPS (Instructions per Second) for maximal throughput [13, 14], but their greedy approach often results in local optimum solutions. While these strategies increase contributions from favored threads, they do not necessarily improve the final performance [2].

In contrast, BCB approaches [2, 15-20] try to fully utilize big cores for better HEMC efficiency. For instance, Koufaty et al. [18] schedule computationally-intensive processes on big cores and memory-intensive ones on small cores to avoid idling on big cores due to slow memory accesses. Van et al. [14] suggest more specifically that ILP and MLP can better quantify big core utilization efficiency than memory and computation intensity. In fact, these ideas are similar to the state-of-the-art IPC-driven approach [15] developed earlier, which schedules processes of higher performance SFs to big cores and the remaining processes to small cores. Essentially, it attempts to minimize the overall workload completion time by leveraging the computation capability of big cores.

In sum, none of these heuristics can guarantee optimum results, and in some cases the results are far from optimum. As shown in the example in the introduction, the IPC-driven approach results in 40% overhead compared to the optimum result, which can be achieved following the methods outlined in the next section.

## 3. EXACT OPTIMUM SOLUTIONS

In this section, we first briefly introduce the scheduling problem and elaborate on our exact analytical optimum solution method.

We define the objective function of the HEMC scheduling problem to be maximizing the throughput or minimizing the total completion time for a given set of workloads on a target system. The key is to consider the differences of process SF values due to different computational/memory patterns.

In practice, each process can have several program phases [9, 19]. Within each process phase, the process exhibits little performance variation. To simplify discussions, we assume each process has one uniform phase and the whole process bears one static SF value.

We also assume negligible cold cache effect caused by cross-core migration and shared cache contentions from co-scheduled processes. In this paper we consider only sequential execution processes and no simultaneous multithreading (SMT). Therefore, at any time each core can execute at most one process and a process can be executed only on one core, not simultaneously on multiple cores.

With these assumptions, the scheduling problem can be defined as minimizing the total completion time for partitioning and executing all workloads/processes.

To keep the formulation concise, we now define a few notations. First, we have $T_k^b$ represent the total execution time of the process $P_k$ on a big core. If $F_k$ is the performance SF of $P_k$, then the total execution time $T_k^s$ on a small core is equal to $T_k^b * F_k$. Note that $F_k$ is normally greater than 1.

We further use $x_k$ to indicate the ratio of $P_k$ executed on a big core, where $x_k$ is of a value between 0 and 1. Thus $P_k$ spends $T_k^b * x_k$ execution time on a big core and $T_k^s * (1 - x_k)$ execution time on a small core. The total execution time of process $P_k$ can then be denoted as

$$t_k = T_k^b * x_k + T_k^s * (1 - x_k) \quad (1)$$

We will later consider a general case where $N$ processes are executed on $B$ big cores and $S$ small cores with $N \geq B + S$. If $N < B + S$, the optimum solution is trivial, as we shall minimize completion time by first putting the longest running processes on big cores and executing the remaining ones on small cores because the longest-running process essentially determines the total completion time.

We shall begin by studying the optimum solution method in the simplest case, running two processes on an HEMC with one big core and one small core, to gain some insight into the problem before gradually extending the solution method to more complicated cases.

A. Running Two Processes on One Big and One Small Core ($1B1S$)

Assume that $P_1$ and $P_2$ are the two processes under consideration. Then, the final total completion time $t^f$ will be determined by the maximal value of the total execution time (1) $t_1$ of $P_1$, (2) $t_2$ of $P_2$, (3) $t^b$ of the workload on the big core, and (4) $t^s$ of the workload on the small core, as shown in the formulated Linear Programming (LP) listed below.

$$\min \quad t^f$$
$$\text{s.t.} \quad t^f \geq t_k = T_k^b * x_k + T_k^s * (1 - x_k), \ k=1, 2$$
$$t^f \geq t^b = T_1^b * x_1 + T_2^b * x_2$$
$$t^f \geq t^s = T_1^s + T_2^s - (T_1^s * x_1 + T_2^s * x_2)$$
$$t^f \geq x_1, x_2 \geq 0$$

Instead of just using the Linear Programing (LP) package to solve the above formulation, we attempt in the following an analytical approach in order to gain insight into the problem.

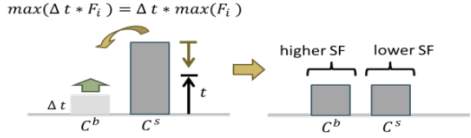

Figure 2: An intuitive optimization approach is to move the process with a higher SF from the small core to the big core until balance is reached

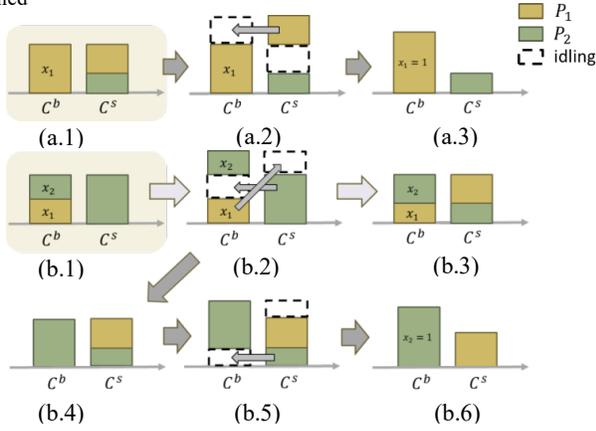

Figure 3: Illustrations of the optimum solutions derived from the balanced intuitive results from Figure 2. (a) A case with overlapping execution of process $P_1$. (b) A case with overlapping execution of process $P_2$.

First, we start from having all two processes on the small core and consider the best way to reduce the total completion time. As shown in Figure 2, if we move $\Delta t$ time units of the process $P_k$ to the big core, it actually reduces $\Delta t * F_k$ time units from the small core. Therefore, to maximize the reduction of the total completion time, we must maximize $\Delta t * F_k$. Since $max(\Delta t * F_k) = \Delta t * max(F_k)$, the best reduction of the total execution time can be achieved by moving the process with a higher SF to the big core. This is the basic idea of the IPC-driven heuristic.

Since ideally no cores should be idling in order to best utilize the computing power, we simply follow the above idea and move processes until the workloads of both cores are balanced. At the end, we shall have one of the two cases shown in Figure 3(a.1) and 3(b.1), under the assumption that $F_1 > F_2$.

For the first case shown in Figure 3(a.1), some of the $P_1$ workload is left in the small core at the point of balance. Since no same-process concurrent execution is allowed, the total completion time is dominated by $t_1$, the total execution time of the process $P_1$, as illustrated in Figure 3(a.2). Clearly, to have the minimum $t_1$, we should execute $P_1$ completely on the big core, as shown in Figure 3(a.3). Hence $t_1 = T_1^b$, which is also the minimum total execution time. Note that in terms of the LP solution, this occurs when $x_1 = 1$, $x_2 = 0$, $t_1 = t^b \geq t_2 = t^s$, or equivalently, when $T_1^b \geq T_2^s$.

Now we discuss the second case shown in Figure 3(b.1), in which all $P_1$ and some of $P_2$ are moved to the big core at the point of balance. In this case, the total execution time is dominated by $t_2$, the total execution time of the process $P_2$, as illustrated in Figure 3(b.2). In this case, part of big-core $P_1$ should be traded with small-core $P_2$ to reduce the idling time while attempting to keep the workloads of both cores balanced. Through some detailed analysis, we find that the determining condition is to check if $T_1^s \geq T_2^b$. If so, we have $t_1 = t^b = t_2 = t^s$, and the workload is balanced. Further computation gives

$$x_1 = (1 - T_2^b/T_1^s)/(1 - 1/F_1 F_2) \quad (2)$$

$$x_2 = (1 - T_1^b/T_2^s)/(1 - 1/F_1 F_2) \quad (3)$$

as shown in Figure 3(b.3).

Otherwise, when $T_2^b \geq T_1^s$, $P_2$ has overlap when the workloads of both cores are balanced, as shown in Figure 3(b.4). In this case, $P_2$ should clearly be completely moved to the big core to minimize total completion time, as shown in Figure 3(b.5) Hence, we finally have $t_2 = t^b \geq t_1 = t^s$, and $x_1 = 0$, $x_2 = 1$, as shown in Figure 3(b.6).

According to the above analysis, we can make the important conclusion that if the big core execution time $T_k^b$ of a process $P_k$ is longer than the execution time of the other processes on the small core, then this process $P_k$ has to be assigned to execute on the big core and the other process assigned to the small core in order to have minimum total completion time or optimum throughput. Otherwise, we can compute $x_1$ and $x_2$ according to Equations (2) and (3) to balance the workload for optimum throughput.

Next, we shall extend the analysis to the cases involving $N$ processes on HEMCs with one big core and one small core.

B. *Running $N$ Processes on One Big and One Small Core*

Assume there are $N$ processes under consideration. Then, the final total execution time $t^f$ will be the maximum value of (1) $t_k$ for each process $P_k$, (2) $t^b$ of the workload on the big core and (3) $t^s$ of the workload on the small core. Accordingly, we formulate the HEMC scheduling problem as an LP problem listed in the following:

$$\min \quad t^f$$
$$\text{s.t.} \quad t^f \geq t_k = T_k^b * x_k + T_k^s * (1 - x_k), \quad k=1 \text{ to } N$$
$$t^f \geq t^b = \sum_{\forall k} T_k^b \cdot x_k$$
$$t^f \geq t^s = \sum_{\forall k} T_k^s \cdot (1 - x_k)$$
$$1 \geq x_k \geq 0, \forall k$$

Again, instead of solving the LP formulation, we try to derive an exact analytical solution based on the initial balanced solution by moving the processes with higher SFs to the big core, as shown in Figure 2. Assume that process $P_k$ is the one that splits into two parts at the balanced situation following the moving sequence, as shown in Figure 4(a).

We now discuss how to partition the workloads at the balanced situation and identify $P_k$ and propose a method for finding the optimum solution. Note that the processes are in descending SF order, i.e. $F_1 \geq \cdots \geq F_k \geq \cdots \geq F_N$. Then, we have $t^b = t^s$ at the balance point, where

$$t^b = \sum_{i=1}^{k-1} T_i^b + x_k T_k^b;$$
$$t^s = (1 - x_k) T_k^s + \sum_{i=k+1}^{N} T_i^s.$$

By solving $t^b = t^s$ for $x_k$, we have

$$x_k = (\sum_{i=k}^{N} T_i^s - \sum_{i=1}^{k-1} T_i^b)/(T_k^b + T_k^s). \quad (4)$$

Since $x_k$ is required to be no less than 0 and no more than 1, equivalently we have

$$\sum_{i=1}^{k-1} T_i^b \leq \sum_{i=k}^{N} T_i^s \ \& \ \sum_{i=1}^{k} T_i^b \geq \sum_{i=k+1}^{N} T_i^s \quad (5)$$

The above equation suggests that to identify $P_k$, we can check sequentially from $P_1$ until we find $P_k$ that satisfies the above constraint.

Consequently, we have $x_k$ by Equation (4) and $x_i = 1$ if $i < k$ or $x_i = 0$ if $i > k$.

If this solution does not have overlaps with $P_k$ or if $t^b = t^s \geq t_k$, as shown in Figure 4(a), then we have an optimum solution; otherwise, we will have to avoid concurrent execution, as illustrated in Figure 4(b.1).

To reduce the idling time of the cores shown in Figure 4(b.2), we can trade processes with lower SFs from the big core with the small-core portion of $P_k$ until either no more processes other than $P_k$ are left in the big core or the workloads of the two cores are balanced with no idling.

We now examine the first case, where no more processes other than $P_k$ are left in the big core and $P_k$ shall be fully assigned to the big core for minimum completion time, as shown in Figure 4(b.3). For this case, the necessary condition is

$$T_k^b \geq \sum_{i \neq k} T_i^s. \tag{6}$$

Otherwise, the optimum solution occurs after moving lower SF processes on the big cores back to the small cores until a process $P_j, j < k$, is partitioned into two parts to balance the workload for optimum solution, i.e., $t^b = t^s = t_k$ as shown in Figure 4(b.4). Then, we have

$$x_k T_k^b = (1 - x_j) T_j^s + \sum_{i=j+1, \neq k}^{N} T_i^s,$$

$$\sum_{i=1}^{j-1} T_i^b + x_j T_j^b = (1 - x_k) T_k^s.$$

Solving the above equations, we have

$$x_j = \left(\sum_{i=1}^{j-1} T_i^b + F_k \sum_{i=j, \neq k}^{N} T_i^s - T_k^s\right) / (F_k T_j^s - T_j^b). \tag{7}$$

Since $x_j$ is required to be between 0 and 1, we derive from the above equation and have

$$\sum_{i=1}^{j-1} T_i^b + F_k \sum_{i=j, \neq k}^{N} T_i^s \geq T_k^s \tag{8}$$

$$\sum_{i=1}^{j} T_i^b + F_k \sum_{i=j+1, \neq k}^{N} T_i^s \leq T_k^s. \tag{9}$$

These two constraints imply that we can sequentially check processes in decreasing order of SFs beginning with $P_{k-1}$ until identifying the process $P_j$. Then, we can use Equation (7) to calculate the optimum solution.

In summary, for the case where $N$ processes are run on $1B1S$, we use Equation (5) to identify the key process $P_k$. If there is no overlap of $P_k$ for the partition based on Equation (4), then we have the optimum solution. If there is overlap, we check the condition of Equation (6) to see if $P_k$ should run on the big core exclusively; otherwise, we check for $P_j$ following the constraints of Equations (8) and (9) and then compute the optimum partition of $P_j$ using Equation (7) and adjust $x_k$ accordingly. The processes from $P_1$ to $P_{j-1}$ are assigned to the big core and those from $P_{j+1}$ to $P_N$, except $P_k$, are assigned to the small core.

For the general case where $N$ processes are being run on $B$ big and $S$ small cores, deriving the exact analytical optimum solution is complicated. We will approach the problem from the process perspective to gain some critical insights and apply them to solve the general case.

C. Process View for Solving Optimum Solutions

Here, instead of stacking the assigned processes on the big/small cores, we use two colors, blue and gold, to indicate the core assignments of each process. Then, the initial balanced results in Figure 4(a) can be mapped equivalently to those in Figure 5(a). The processes with higher SFs than that of $P_k$ are assigned to a big core, as indicated in blue, and the rest in gold are assigned to a small core, while $P_k$ is split into two parts in order to keep a balanced result, i.e., $t^b = t^s$.

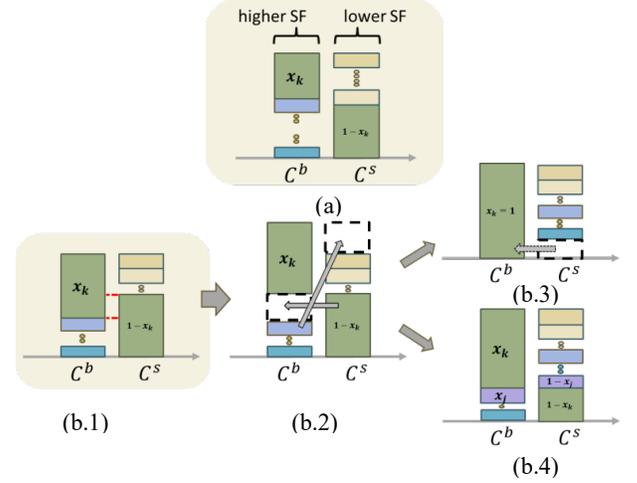

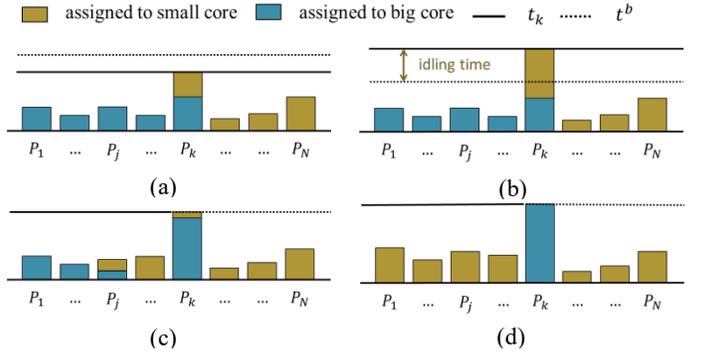

Figure 4: Illustrations of optimum solutions derived from the balanced intuitive results from Figure 2. (a) A case with no same-process concurrent execution (b) A case with overlapping execution of process $P_k$

Figure 5: Illustrations of the optimum solution algorithm from a process perspective. (a) The equivalent solution of Figure 4(a); (b) the equivalent solution of Figure 4(b.1); (c) the equivalent solution of Figure 4(b.4); (d) the equivalent solution of Figure 4(b.3);

To make the results visually comprehensive, we use a solid line to indicate the $t_k$ of $P_k$ and a dotted line to mark the $t^b$ of the workload on the big core. In cases where the dotted line is no lower than the solid line, we have $t^b \geq t_k$ and the result is optimum, as shown in Figure 5(a), which corresponds to Figure 4(a).

Otherwise, $t_k > t^b$ and the solid line is higher than the dotted line, as shown in Figure 5(b). In this case, the idling time is simply the gap between $t_k$ (solid line) and $t^b$ (dotted line). To reduce idling time, we can trade processes of lower SFs in the big core with the small-core portion of $P_k$.

Now assume we pick a candidate process $P_c$ to trade in the order of $P_{k-1}$ to $P_1$. If the small core portion of $P_k$ is sufficient for trading with $P_c$, we then move $P_c$ to execute on the small core and reassign part of the small-core portion of $P_k$ to the big core while maintaining workload balance, i.e., $t^b = t^s$. Otherwise, only part of $P_c$ (or $P_j$ in Figure 5(c)) can be traded with $P_k$, and we then compute the exchange that balances the workload ($t^b = t^s$) to arrive at the balanced result, as shown in Figure 5(c).

Note that through this procedure, we are essentially reducing the gap between the solid line ($t_k$) and the dotted line ($t^b$) by making one lower and the other higher until idling time is eliminated.

However, if we trade all candidate processes and still have a small-core portion of $P_k$ remaining, then we can only close the gap further and reduce idling time by moving the entire remaining small-core portion of $P_k$ to the big core. Consequently, we have unbalanced results, i.e., $t_k = t^b > t^s$, as shown in Figure 5(d), which corresponds to Figure 4(b.3).

Using the process view above, we can derive the same optimum solution. Although this seems to be redundant, we find that the process view is more appropriate for solving the general case of running $N$ processes on $B$ big and $S$ small cores. Basically, we find that optimum solutions can be derived using greedy reduction of the execution time of the longest processes of the initial balanced result through trading.

D. Running $N$ Processes on $B$ Big and $S$ Small Cores

Now assume there are $N$ processes and $B$ big and $S$ small cores under consideration. Since the ideal optimum solution occurs when it is balanced, i.e., $t^b/B = t^s/S$, as shown in Figure 6, we may simply extend the previous algorithm in Sec. 3.C to this scenario. We will show later that the final total completion time $t^f$ is the maximum of (1) $t_{max}$ of the longest running processes, (2) $t^b/B$ of average big core workload, and (3) $t^s/S$ of average small core workload. In other words, $t^f = max\{t_{max}, t^b/B, t^s/S\}$.

As shown in Figure 7, we have two possibilities for the initial solution with $t^b/B = t^s/S$: one has $t^b/B = t^s/S \geq t_{max}$ and the second has $t^b/B = t^s/S < t_{max}$. In the case where $t^b/B = t^s/S \geq t_{max}$, shown in Figure 7(a), the scenario is similar to what we had in Figure 4(a), and the result is an optimum with a total completion time equal to $t^b/B$. The case shown in Figure 7(b), where $t^b/B = t^s/S < t_{max}$, can be further divided into two scenarios: either the corresponding process $P_{max}$ of $t_{max}$ is completely assigned to big cores, as shown in Figure 8(a), or it is partially or completely assigned to small cores, as shown in Figure 8(b). Note that in Figure 8, blue indicates assignment to a big core and gold to a small core. Additionally, we use the solid line to indicate $t_{max}$ of the longest running process $P_{max}$ and the dotted line to indicate $t^b/B$ of the average big core workload.

Now for the scenario in Figure 8(a), since the process $P_{max}$ is all assigned to a big core and its execution time $t_{max}$ can be reduced no further, we have an optimum result.

Otherwise, for the scenario in Figure 8(b), we can reduce idling time by following a procedure similar to the 1B1S case by iteratively trading the small core portion of $P_{max}$ with the candidate process $P_c$, the smallest SF process that is either partially or completely assigned to big cores. Note that for this scenario, $P_{max}$ can happen to be $P_k$.

With each trade, the value of $t_{max}$ will gradually become lower while the big and small core workloads maintain balanced, i.e., $t^b/B = t^s/S$ and $t_{max}$ is lessened. Then, we have two possible results: either $P_c$ is fully or partially traded.

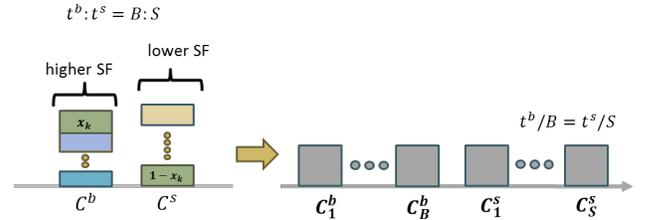

Figure 6: The intuitive balanced solutions after moving processes of higher SFs from the small core pool to big core pool

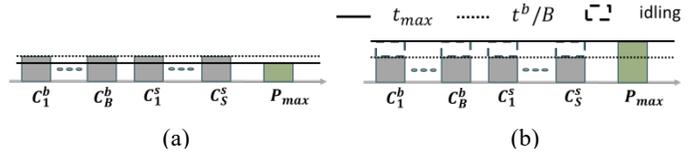

Figure 7: Illustrations of the initial balanced results from Figure 6. (a) The case where $t^b/B = t^s/S \geq t_{max}$; (b) the case where $t^b/B = t^s/S < t_{max}$

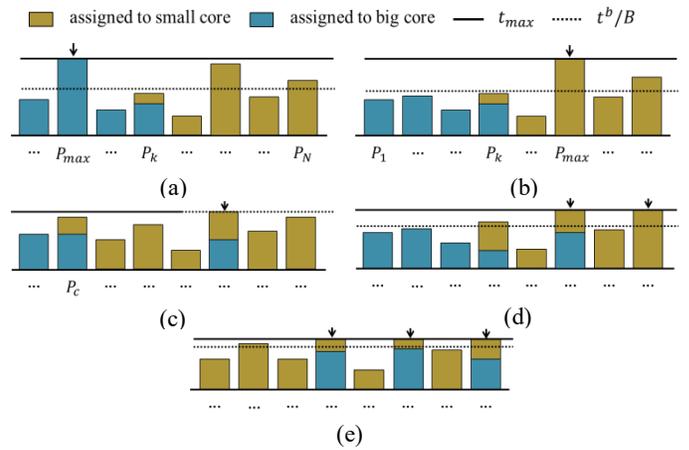

Figure 8: Illustrations of the optimum solution algorithm from the process perspective. (a) The longest running process is fully assigned to a big core; (b) the longest running process, or processes, are partially assigned to a small core; (c) the optimum result when $t^b/B = t_{max}$; (d) the longest running process is lowered to match the second longest running process; (e) the longest running processes are lowered when all candidate processes are traded to small cores by moving as much as possible to big cores.

If $P_c$ is fully traded, we continue to find next candidate process for trade; otherwise, if $P_c$ is partially traded, we shall have the situation shown in Figure 8(c), where the execution time of the longest running process coincides with that of the balanced workload of the cores. This is an optimum solution, as we can no longer make improvement.

However, if the originally longest running processes are lowered to match the second longest running process, as illustrated in Figure 8(d), we then have a new set of longest running processes. If any of the longest running processes is fully assigned to big cores, as shown in Figure 8(a), then we have optimized the solution and cannot improve it further; otherwise, we can repeat the above trading procedure to lower this new set of longest running processes.

In a case where all the processes with higher SFs than that of the longest running processes have already been fully traded to small cores, as shown in Figure 8(e), we cannot find any process candidates (similar to Figure 5(d)) for improvement. For this case, in order to further reduce

$t_{max}$ we may simply move as many of the small-core portions of the longest running processes to big cores as possible, producing an unbalanced optimum result such that $t_{max} \geq t^b/B > t^s/S$.

We conclude from the discussions above that the optimum solution must be one of the three cases shown in Figure 8(a),(c),(e) and the final total completion time $t^f$ is equal to either $t_{max}$ or $t^b/B$ (or $t^s/S$ when $t^b/B = t^s/S \geq t_{max}$). In general, there are one or more longest running processes whose total execution time is exactly $t^f$. Note that only these longest running processes and at most one more process, i.e., $P_c$ in Figure 8(c) whose execution time is less than $t^f$, can have split assignments to both big and small cores. The rest are fully assigned to either big or small cores.

For completeness, we now discuss how to realize the big/small core scheduling assignments under the computed optimum total completion time. We shall use the bin packing method for scheduling. Basically, each core is represented as a bin with the size $t^f$ and the process packed to the bottom is scheduled to execute first. Since the same process cannot be executed concurrently, there shall be no horizontal overlap of the assigned sections of any process. In principle, we naturally avoid overlap by trying to assign the big core portion of a process from bottom up and the small core portion from top down.

We now first assign the group of longest running processes, as illustrated in Figure 9. For the very first longest running process (colored purple), since there are no other occupants, we simply place the big core portion at the very bottom and the small core portion at the very top, as shown in Figure 9(a.1). If we fold the assignments into one bin, we find the two assignments have no overlap and fill up the whole bin with no gap: this is illustrated in Figure 9(a.2), where the big core portion is blue and the small core portion is gold.

Figure 9(b.1) illustrates what happens if there is another longest running process (colored light blue). Here, we essentially assign the big core portion from bottom up following the end point of the previous big core assignment, marked as the dotted line in Figure 9(a.1), and the small core portion top down from the end of the previous small core assignment. In case the bin under assignment runs out of space, we continue with the rest from the very bottom or very top of a new bin, depending on whether it is a big or small core assignment. The folded view shown in Figure 9(b.2), illustrates that there is no gap and no collision between the big core (blue) portion assigned bottom up from the dotted line of Figure 9(a.1) and the small core (gold) portion. Hence, by repeating the same procedure on the remaining longest running processes, we can produce a legal assignment without collision.

Now if there exists a process $P_c$ (colored gray green in Figure 9(c.1)) with a split big/small core assignment and execution time smaller than $t^f$, we essentially follow the same assignment procedure as for the longest running process, except that at the end there will be a gap remaining even though the assignment is guaranteed to have no overlap, as shown in Figure 9(c.2).

Finally, the remaining processes are either fully assigned to big cores or small cores. We subsequently schedule the big core processes from bottom up following the end point of the last big core assignment and the small core processes top down from the end of the last small core assignment. If the bin under assignment runs out of space, we continue the rest from the very bottom or top of a new bin depending on whether it is a big or small core assignment. This procedure is illustrated in Figure 9(d). After all processes are scheduled, we reach the final solution.

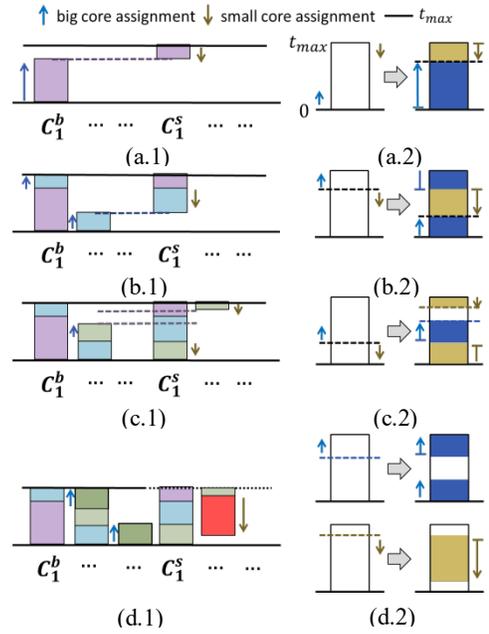

Figure 9: Illustration of the process scheduling assignment steps. (a) Scheduling of the first longest running process; (b) scheduling of the subsequent longest running processes; (c) scheduling of the non-longest process with split big/small core assignment; (d) scheduling of the remaining full big/small core assigned processes.

The bin capacity is guaranteed to be sufficient, as the big/small core assignments are computed using the optimum algorithm. Since our bin-packing-based scheduling method wastes no idling space in between processes, the end result is guaranteed to be legal and feasible.

With the above scheduling procedure, we complete our optimum solution algorithm.

Next, we analyze the time complexity of the proposed algorithm.

E. Time Complexity

Our proposed optimum solution algorithm for the general case in Section 3.D consists of three steps. We first generate an initial balanced result, then repeatedly perform trading to reduce idling time, and finally move excess small core portions of the longest running processes to the big cores if necessary. Essentially, the second step dominates the complexity.

In step one, we basically scan through the ordered processes to find an initial balanced workload solution. The sorting takes O(*N*log*N*) time and the scanning is linear time. Step three requires only a simple calculation to find a proper amount to move over, so the complexity equals that of the trade iteration in second step. We analyze this step, which is the most complicated, next.

For the second step of the optimization algorithm, we have the cases in Figures 8(a), 8(c) and 8(e) indicating the three possible end results of the algorithm. The computational complexity resides in the iterative procedure in Figures 8(b) and 8(d) that tries to match the second longest running process. Since the number of longest running processes cannot be greater than the number of cores *C*, where *C*=*B*+*S*, we have at most *C* iterations. Throughout the iterations, we may also scan through all the ordered big core processes for trading and this clearly is at most *N* times.

Yet, if we suppose that there are *k* longest running processes for each iteration, we must scan through these *k* processes to determine the proper amount of small core portions to trade, which requires *k* computations. To maintain the information of the second longest process, we apply the red-block tree data structure [21], which requires O(*N*) to initialize and O(log *N*) to maintain at each iteration. Therefore, the complexity is O((*N* + *C*)(*C* + log *N*)).

After summing up the complexity numbers of the three steps, we have the total complexity of $O(NC + C^2 + (N + C)\log N)$. For a fixed HEMC, *C* is constant and hence the time complexity is O(*N* log *N*). The complexity analysis shows that our algorithm is efficient for practical use.

## 4. EXPERIMENTAL RESULTS

### A. Experimental Environment Setup

For the testing, we implement our optimum algorithm and other reference algorithms on Sniper [22], a hardware-validated simulation tool, mainly because it supports HEMC architecture. In Table 2 we show the specification of the big/small cores. The big cores have out-of-order (OOO) pipeline execution with reorder buffers (ROBs) and can issue multiple instructions whereas the small cores have in-order (IO) pipeline execution with lower ILP and MLP. Both core types run at a 2.66 GHz clock rate and have the same cache configuration, a 3-level private cache hierarchy with 64 KB (32 for I-cache and 32 for D-cache), 256 KB, and 8 MB caches. We then configure target HEMCs so each can be tested with a total number of *C* cores ranging from 2 to 10 cores, with the big core number *B* varying from 1 to *C-1*.

For our test cases, we adopt the popular SPEC CPU 2006 benchmark suite, among which 19 test cases are accepted by Sniper, including nine float-point processes and ten integer ones. With Sniper, we first profile each test, process the big core and small core performance values, and calculate the corresponding SF value. In Figure 10, we show these 19 cases in descending SF order and the corresponding big/small core run times. Note that SF values of test cases vary from 1.09 to 2.5 and some feature longer run times compared to others.

In the following, we show compare the performance of our proposed optimum (OPT) approach with the IPC-driven (IPC) approach [15] and the long-time-first (LTF) approach, which prioritizes placement of long processes on big cores.

### B. Performance Evaluation

We show the comparison results in Figure 11, in which the execution times are all normalized in reference to our generated optimum results. Note that the figure is a 2-D representation from a 3-D graph with the primary index (printed number) being the total number *C* of all cores, and the secondary index (implicit) being the number *B* of big cores.

First, we pool together a copy of each SPEC benchmark process for testing. The comparison results shown in Figure 11(a) indicate that the IPC approach incurs an overhead over the optimum approach of 18.7% on average and 33.9% in the worst case. In contrast, the LTF approach has only an average overhead of 7.5% and worst case overhead of 10.8%. The reason why LTF performs better than IPC in this test is due to the fact that some processes are relatively long and the final completion time is dominated by these long processes, which are better to execution on big cores.

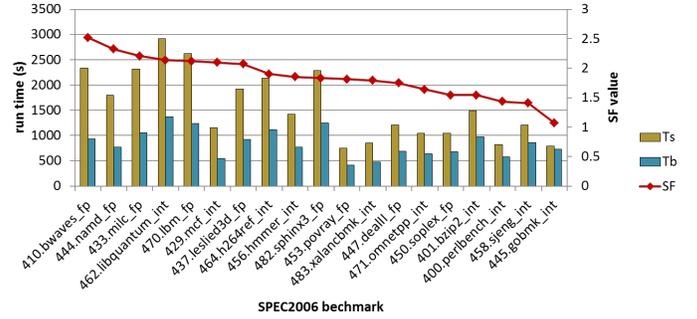

Table 2: Core configurations

| Core Configuration | Big core (OOO) | Small core (IO) |
|---|---|---|
| Dispatch Width | 4 | 1 |
| ROB size | 36 | X |
| Cache size(L1, L2, L3) | 32+32 KB, 256KB, 8MB | |

Figure 10: The execution time on a big/small core of 19 benchmarks chosen from the SPEC2006 suite. They in descending order by SF value.

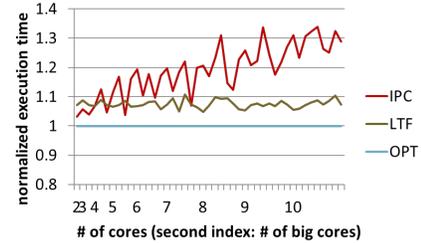

(a) Test results on a mix of one copy of each SPEC benchmark case.

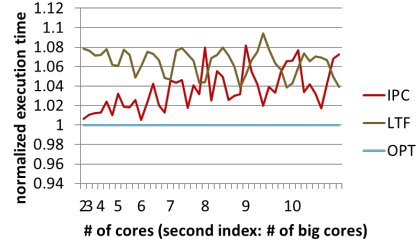

(b) Test results on a mix of 5 copies of each SPEC benchmark case.

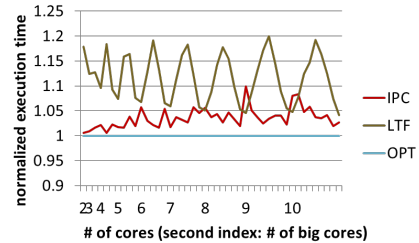

(c) Test results on a mix of 40 copies of *bwaves* and *gobmmk*.

Figure 11: Comparing the normalized execution time results of IPC and LTF with our proposed optimum method.

For the second test, we mix 5 copies of each SPEC process. The intention is to dilute the influence of the long processes. As the results show in Figure 11(b), we observe that for this case the IPC approach performs close to the optimum one, with an average overhead of 3.6%. In contrast, the LTF approach does not effectively utilize big cores and the average overhead is about 6.4%.

Finally, to stress test the LTF approach, we make a new test case with 40 copies of the process *bwaves*, whose SF is 2.5, and 40 copies of the *gobmmk* process, whose SF is only 1.09. As the results show in Figure 11(c), the IPC approach performs well, with an average overhead of 3.6%, while the LTF approach suffers an average overhead of 11.5% and worst case overhead of 19.9% due to acute loss of big core utilization efficiency.

The above evaluations confirm that our implementation does produce optimum results. More importantly, our proposed optimum algorithm is robust, unlike the IPC and LTF approaches, which may perform worse under certain conditions.

## 5. CONCLUSIONS AND FUTURE WORK

In this paper, we have proposed an analytical optimization method for the HEMC process scheduling problem. Our contribution an efficient optimum solution method developed after precisely analyzing the optimization problem. The approach is robust and avoids issues of existing algorithms, which are either biased toward long or high-SF processes.

In the future, we plan to extend our work to consider program phase behaviors and cache contention effects among co-executing processes. Additionally, we also hope to devise optimization methods for other objective functions, such as energy efficiency, performance/power efficiency, or service quality fairness awareness.

## 6. REFERENCES

[1] Kumar, Rakesh, et al. "Single-ISA heterogeneous multi-core architectures: The potential for processor power reduction." Microarchitecture, 2003. MICRO-36. Proceedings. 36th Annual IEEE/ACM International Symposium on. IEEE, 2003.

[2] Kumar, Rakesh, et al. "Single-ISA heterogeneous multi-core architectures for multithreaded workload performance." ACM SIGARCH Computer Architecture News. Vol. 32. No. 2. IEEE Computer Society, 2004.

[3] AMD. The future is fusion: The industry changing impact of accelerated computing. http://sites.amd.com/jp/Documents/AMD_fusion_Whitepaper.pdf, 2008

[4] NVidia. Variable SMP – a multi-core CPU architecture for low power and high performance. http://www.nvidia.com/content/PDF/tegra_white_papers/Variable-SMP-A-Multi-Core-CPU-Architecture-for-Low-Power-and-High-Performance-v1.1.pdf, 2011

[5] OpenCL Specification v2.0, Khronos Group, Oct. 2009. [Online]. Available: http://www.khronos.org/registry/cl

[6] Auerbach, Joshua, et al. "A compiler and runtime for heterogeneous computing." Proceedings of the 49th Annual Design Automation Conference. ACM, 2012.

[7] P. Greenhalgh. Big.LITTLE processing with ARM Cortex-A15 & Cortex-A7: Improving energy efficiency in high-performance mobile platforms. http://www.arm.com/files/downloads/big_LITTLE_Final_Final.pdf, Sept. 2011.

[8] Van Craeynest, Kenzo, et al. "Fairness-aware scheduling on single-ISA heterogeneous multi-cores." Parallel Architectures and Compilation Techniques (PACT), 2013 22nd International Conference on. IEEE, 2013.

[9] Sawalha, Lina, and Ronald D. Barnes. "Energy-efficient phase-aware scheduling for heterogeneous multicore processors." Green Technologies Conference, 2012 IEEE. IEEE, 2012.

[10] Cong, Jason, and Bo Yuan. "Energy-efficient scheduling on heterogeneous multi-core architectures." Proceedings of the 2012 ACM/IEEE international symposium on Low power electronics and design. ACM, 2012.

[11] Isci, Canturk, et al. "An analysis of efficient multi-core global power management policies: Maximizing performance for a given power budget." Proceedings of the 39th annual IEEE/ACM international symposium on microarchitecture. IEEE Computer Society, 2006.

[12] Teodorescu, Radu, and Josep Torrellas. "Variation-aware application scheduling and power management for chip multiprocessors." ACM SIGARCH Computer Architecture News. Vol. 36. No. 3. IEEE Computer Society, 2008.

[13] Liu, Guangshuo, Jinpyo Park, and Diana Marculescu. "Dynamic thread mapping for high-performance, power-efficient heterogeneous many-core systems." Computer Design (ICCD), 2013 IEEE 31st International Conference on. IEEE, 2013.

[14] Van Craeynest, Kenzo, et al. "Scheduling heterogeneous multi-cores through performance impact estimation (PIE)." ACM SIGARCH Computer Architecture News 40.3 (2012): 213-224.

[15] Becchi, Michela, and Patrick Crowley. "Dynamic thread assignment on heterogeneous multiprocessor architectures." Proceedings of the 3rd conference on Computing frontiers. ACM, 2006.

[16] Sondag, Tyler, Viswanath Krishnamurthy, and Hridesh Rajan. "Predictive thread-to-core assignment on a heterogeneous multi-core processor." Proceedings of the 4th workshop on Programming languages and operating systems. ACM, 2007.

[17] Saez, Juan Carlos, et al. "Leveraging workload diversity through OS scheduling to maximize performance on single-ISA heterogeneous multicore systems." Journal of Parallel and Distributed Computing 71.1 (2011): 114-131.

[18] Koufaty, David, Dheeraj Reddy, and Scott Hahn. "Bias scheduling in heterogeneous multi-core architectures." Proceedings of the 5th European conference on Computer systems. ACM, 2010.

[19] Sawalha, Lina, et al. "Phase-guided scheduling on single-ISA heterogeneous multicore processors." Digital System Design (DSD), 2011 14th Euromicro Conference on. IEEE, 2011.

[20] Lugini, Luca, Vinicius Petrucci, and Daniel Mosse. "Online thread assignment for heterogeneous multicore systems." Parallel Processing Workshops (ICPPW), 2012 41st International Conference on. IEEE, 2012.

[21] Bayer, Rudolf. "Symmetric binary B-trees: Data structure and maintenance algorithms." Acta informatica 1.4 (1972): 290-306.

[22] Carlson, Trevor E., Wim Heirman, and Lieven Eeckhout. "Sniper: exploring the level of abstraction for scalable and accurate parallel multi-core simulation." Proceedings of 2011 International Conference for High Performance Computing, Networking, Storage and Analysis. ACM, 2011.